\documentclass[12pt]{article}
\usepackage{times,latexsym}
\usepackage{amsmath}
\usepackage{amssymb}
\usepackage{stmaryrd}
\usepackage{graphicx}
\usepackage{epsfig}
\usepackage{lineno}
\usepackage{fancyhdr}
\usepackage[usenames,dvipsnames]{color}

\usepackage[square,comma,numbers,sort&compress]{natbib}

\begin{document} 


\bibliographystyle{unsrtnat}

\centerline{\Large{Astro2020 Decadal Science White Paper:}}
\centerline{\Large{The state of gravitational-wave astrophysics in 2020}}

\vspace{0.5in}

\noindent{\bf Thematic Areas:}\\
$\square$ Planetary Systems \\
$\square$ Star and Planet Formation \\
$\square$ Formation and Evolution of Compact Objects \\ 
$\boxtimes$ Cosmology and Fundamental Physics \\
$\square$  Stars and Stellar Evolution \\ 
$\square$ Resolved Stellar Populations and their Environments \\
$\boxtimes$ Galaxy Evolution \\ 
$\boxtimes$ Multi-Messenger Astronomy and Astrophysics

\vspace{0.5in}

\noindent \textbf{Principal Author:} \\
\noindent Sean T. McWilliams,\\
\noindent West Virginia University,\\
\noindent {\tt Sean.McWilliams@mail.wvu.edu}\\
\noindent (304) 435-5311

\vspace{0.5in}

\noindent \textbf{Co-authors:} \\
\noindent Robert Caldwell, Dartmouth College,\\
\noindent Kelly Holley-Bockelmann, Vanderbilt University,\\
\noindent Shane L. Larson, Northwestern University,\\
\noindent Michele Vallisneri, Jet Propulsion Laboratory

\vspace{0.5in}

\noindent \textbf{Abstract:} \\
\noindent While still in its infancy, the budding field of gravitational-wave astronomy has so far exceeded most expectations, and the achievements that have already been made bode well for the decade to come. While the discoveries made possible by LIGO have captured the imagination of experts and nonexperts alike, it is important when looking ahead to consider those discoveries in the context of the field as a whole.  Just as radio, optical, and x-ray radiation probe different physical phenomena occurring on a range of length and energy scales, the future of gravitational-wave astrophysics depends on our ability to open up the entire spectrum. We will describe the scientific prospects for the field of gravitational-wave astronomy as a whole as we enter the coming decade, and we will place the specific contributions from a future space-based gravitational-wave observatory within this context.

\pagebreak

\section*{Introduction:}

The field of gravitational-wave (GW) astrophysics has witnessed tremendous progress in the past few years, starting with the extraordinary successes of Advanced LIGO in making the first direct detection in 2015 \cite{Abbott:2016blz}, followed by the first observation of gravitational waves and light from a common source by Advanced LIGO (hereafter “LIGO”) and Advanced Virgo in 2017 \cite{TheLIGOScientific:2017qsa,GBM:2017lvd}.  At the same time, both Cosmic Microwave Background (CMB) experiments and Pulsar Timing Arrays (PTAs) are closing in on sensitivity levels where they may reasonably expect to see a signal. In addition, the success of LISA Pathfinder in exceeding its own performance target in 2016 \cite{LPF16}, and ultimately exceeding the final LISA performance target in 2018 \cite{LPF18} has solidified confidence in the viability of space-based gravitational-wave observation.  As a result of these and other developments, the landscape for gravitational-wave science is very different than it was a decade ago. 

\section{Ground-based facilities}

The subfield of ground-based laser interferometers, which includes LIGO and its planned upgrades and future proposed facilities like Voyager, Cosmic Explorer, and the Einstein Telescope \cite{LSCInstrument}, probe gravitational wave frequencies of Hz-kHz.  The principal phenomena expected in this band are the inspirals and merger-ringdowns of binaries containing neutron stars and stellar- and intermediate-mass black holes.  The results from LIGO to date suggest that black-hole binaries are and will continue to be the most common observed source, because although they are intrinsically more rare, their larger masses compared to neutron stars make them visible to greater distances.  During the most recent observing run, the LIGO network of detectors had sufficient sensitivity to observe stellar black-hole binaries out to distances of several Gpc, and to observe neutron-star binaries out to hundreds of Mpc \cite{catalog}.  

In the near future, the current Advanced LIGO and Advanced Virgo facilities are poised to reach their design sensitivities, with additional facilities having comparable sensitivities slated to join the worldwide network over the next several years. In addition, funding has already been approved to begin the next phase of facility upgrades. As early as 2024, these upgrades to the LIGO instruments, dubbed ``A$+$'', will have the capability to observe even farther: approaching 1 Gpc for neutron-star binaries and beyond a redshift of 1 for stellar-mass black-hole binaries like those that have already been observed \cite{ligoaplus}. Additional proposed upgrades to the existing facilities could allow the sensitivity to reach the quantum noise limit down to 5 Hz, which would allow the observation of stellar- and intermediate-mass black-hole binaries out to a redshift of 6, and to localize binaries containing neutron stars to a far smaller area on the sky, which will greatly enhance the prospects for observing electromagnetic counterparts \cite{ligolf}.

In addition to informing our understanding of the distributions of black holes and neutron stars out to those distances, we can infer their environments and formation mechanisms based on their measured masses, spins, and orbital eccentricities.  Observing neutron stars will eventually give us insights into their equation of state, through its effect on the tidal dissipation of energy during the inspiral, and the disruption of the star in the presence of a small enough black hole’s tidal field.  Observing black holes in such a strong-field regime will provide ever-improving tests of general relativity, and may eventually uncover signs of deviations from general relativistic predictions that are indicative of new physics.

One of the highlights in this emerging field over the last few years has been the observation of a complete spectrum of electromagnetic emission \cite{mma} coinciding with the first (and as of 2019, only) detection of a neutron-star binary \cite{bns}.  This event was closer and therefore louder in gravitational waves than was generally expected for this first event, which made it possible to observe its radio and optical emission, and yet it was also visible in x- and gamma-rays, despite being far closer than any other observed gamma-ray burst before or since \cite{grb}.  With the combined information from both types of radiation, we were able to confirm that neutron-star mergers are the progenitor of short gamma-ray-bursts, and that these events are primarily responsible for creating most of the heavy elements in our Universe.  This event was also used to constrain the Hubble constant in a new and systematically different way \cite{hubble}; this method, when applied to more sources in the future, will eventually help clarify the existing disparity between measurements of the Universe$^\prime$s expansion made using data from Planck, and measurements that rely on the use of supernovae as standard candles.

\section{CMB polarization}


While LIGO has made the only direct observation of gravitational waves so far, advances in other wavelengths, using very different methods from LIGO, have reached levels of sensitivity where it is not unreasonable to expect a signal, and the continued absence of a detectable signal places interesting limits on otherwise viable theories of the early Universe \cite{Akrami:2018odb,Ade:2018gkx}.  Gravitational waves with cosmological-scale wavelengths leave a unique pattern, referred to as “B modes”, in the anisotropy pattern of the CMB polarization \cite{Kamionkowski:2015yta}. Detection of this primordial signal would have profound implications for the theory of the early Universe. Sensitivity to B-mode polarization in the CMB has dramatically improved in recent years, and further improvements are underway. These measurements probe the amount of energy in the very early Universe that was contained in gravitational waves, and therefore provide a snapshot of the state of the Universe as it exited its inflationary phase. These waves redshift with the cosmic expansion, so that the CMB signal is dominated by wavelengths on the scale of the cosmological horizon at the time of last scattering and again at reionization. These timescales translate into large angular scales in the present day.

There are a number of CMB polarization experiments pursuing the large-angle B-mode signal, looking for signs of a primordial background of gravitational waves. Current ground-based efforts include BICEP3 \cite{Ahmed:2014ixy}, CLASS \cite{Essinger-Hileman:2014pja}, Polar Bear and the Simons Array \cite{Arnold:2014qym}. Large experiments are also getting underway, including the Simons Observatory \cite{Ade:2018sbj} and the CMB-S4 collaboration \cite{Abazajian:2016yjj}. Balloon-borne experiments include Spider \cite{Gualtieri:2017zcz} and the proposed PIPER \cite{Gandilo:2016sqn}, and satellite missions that are under study include LiteBird (JAXA) \cite{Suzuki:2018cuy} and PICO (NASA) \cite{Young:2018aby}.

\section{Pulsar timing arrays}

At frequencies from 1-100 nHz, three different pulsar timing arrays (PTAs), the European PTA, North American Nanohertz Gravitational Wave Observatory (NANOGrav), and the Parkes PTA, have all reached sensitivities that have begun to constrain theoretical predictions to varying degrees.  The primary target of these observatories is a stochastic broadband signal resulting from the combination of signals from the billions of tightly bound supermassive black-hole binaries throughout the Universe, with the dominant contribution expected to come from binaries at and above a billion solar masses at $z \lesssim 1$.  These binaries are evolving very slowly and are nearly monochromatic over the duration of current data sets.  

The NANOGrav \cite{NANOGrav} and Parkes \cite{Parkes} PTAs have independently published limits on a stochastic signal that are inconsistent with theoretical expectations for the case of purely gravitationally interacting binaries whose masses are assumed from empirical correlations between black hole and galaxy properties that assume relatively large black hole masses.  Most recently, NANOGrav limits have reached the levels where uncertainties in the solar system ephemeris had become a limiting uncertainty, such that the data must now be used to simultaneously account for errors in the ephemeris in addition to any potential correlated signal such as would be caused by a gravitational-wave signal \cite{nano}.

Over the next few years, additional data will confirm or refute the presence of a signal.  Initially, detection evidence will primarily come from a very narrow band, but over time, additional frequencies will become detectable, allowing us to measure the spectrum of the stochastic signal.  The shape and variance of this spectrum will contain information about the mechanism driving binary inspiral at these frequencies, such as gravitational radiation or gas and stellar interactions, as well as information about the distribution of binary properties that contribute most to the signal \cite{nanospec}.  In addition to the stochastic contribution that is expected to be the first observable signal in the data, PTAs may eventually observe several other classes of source. PTAs should be able to identify individual signals from the most massive and closest binaries \cite{nanocw}, and if a massive and nearby binary were to actually merge, PTAs could potentially observe a counterpart to the merger signal in the form of a nonlinear memory \cite{nanobwm}, whereby the higher-frequency gravitational waves sourced by the merger themselves source a gravitational-wave signature at lower frequencies, which is potentially visible in the PTA band.  Apart from signals driven by supermassive black-hole binaries, more exotic sources like cosmic strings and primordial stochastic backgrounds predicted by some theories beyond the standard model could be observed \cite{nano}.

\section{Space-based facilities}

The frequency band above PTAs and below LIGO is the domain of space-based observatories.  PTA sensitivity unavoidably degrades at high frequencies, since white noise in the timing residuals, which are their actual observable, result in a strain sensitivity that strongly favors low frequencies.  Similarly, the seismic wall presents a fundamental limit at low frequencies for LIGO and future upgrades, where advanced seismic isolation techniques can push seismic noise below other noise sources over a modest range of frequencies, but no proposed technology can suppress the noise below a few Hz.  Filling as much of the band between 100 nHz and 1 Hz as possible requires a broadband instrument, and avoiding a seismic wall requires going into space. This reality has been understood for some time, and motivated the conception of a space-based laser interferometer.  However, even demonstrating that an instrument can be adequately isolated from any other disturbances in the absence of seismic noise, requires that you fly an instrument in space. 

The need for this technology demonstration was finally satisfied with the flight of LISA Pathfinder, which succeeded beyond even its designers' wildest expectations.  Pathfinder’s design requirements on strain sensitivity were an order of magnitude less stringent than the final LISA requirement, and no requirements were made below 1 mHz.  However, not only did Pathfinder exceed its own requirements, but it has exceeded the LISA design requirements as well, and has done so not only above 1 mHz, but across the entire frequency band covered by LISA.

With the tremendous success of LISA Pathfinder, the technology required to build LISA is at an advanced state of readiness for a large mission planned for launch in 2034.  In the previous decade, the Astro2010 decadal survey recommended LISA as a priority among L-class missions, but made that recommendation contingent on a successful demonstration of disturbance reduction in space by Pathfinder.  Multiple times, in previous decadal surveys as well as internal NASA and ESA competitions such as the Beyond Einstein Program Assessment Committee, the lack of a successful technology demonstration by Pathfinder was raised as the single greatest risk facing LISA.  With the success of Pathfinder, this final concern has not only been addressed, but has been completely put to rest, paving the way for the full LISA mission.  

Europe has continued to make progress, not only through their leadership on Pathfinder, but by advancing LISA as one of its own L-class priorities.  In 2013, ESA identified the “gravitational universe” as its third science theme, and in June of 2017, the LISA mission concept was selected to fulfill this science theme.  Meanwhile, in the U.S., the mid-decadal review of Astro2010 released its report in 2016 \cite{NWNH}, just after LIGO's discovery of gravitational waves and LISA Pathfinder initial demonstration that it had exceeded its design requirements.  Citing these successes, the review panel advised NASA to restore full support for LISA in the U.S., so that the U.S. community could “be a strong technical and scientific partner”.  A restoration of NASA's role in LISA, even as a minority partner, is essential to providing the U.S. community with opportunities to make technical contributions and gain insights on design and operation, and to facilitate necessary contributions from the U.S. community.

The science expected from LISA bridges the gap between PTAs and LIGO in a number of ways, which will be detailed extensively in other white papers.  Briefly, among its principal science targets, LISA can observe the inspirals and mergers of massive black-hole binaries throughout the visible Universe.  Outside the range between $10^4$--$10^8$ solar masses, LISA can still observe black-hole binaries to substantial distances.  Specifically, LISA could observe the merger of any supermassive binary that would be visible as a resolvable binary or a nonlinear memory event by PTAs.  Both through its observation of more distant, lower-mass sources as well as the most massive nearby sources, LISA will provide an ideal complement to the science provided by PTAs.  Within the paradigm of hierarchical structure formation, the massive nearby PTA sources were assembled through some combination of accretion and mergers by more distant, lower-mass LISA sources.  In terms of the life cycle of a very massive binary, PTAs provide information about the early inspiral of the nearby population, and LISA observes the late inspiral and final merger of the more distant massive population, as well as any rare nearby massive sources that merge. For the most nearby sources, this final merger itself drives a nonlinear memory signal that could be observable by PTAs.  While the event rates for nonlinear memory are not very encouraging, LISA is expected to observe hundreds of mergers, most or all of which will be far, far louder, relative to instrument noise, than can be achieved by ground-based observatories.  As such, LISA presents the best opportunity, by far, to probe the strong-field, nonlinear domain of highly dynamical gravity \cite{BertiWP,ColpiWP}.

For masses below $10^4$ solar masses, LISA can still observe binaries to substantial distances.  For instance, GW150914, Advanced LIGO's first directly detected source, would have been visible earlier in its inspiral by LISA \cite{Sesana:2016ljz}.  This complementarity between LISA and any contemporary ground-based observatory presents a number of exciting scientific opportunities.  For instance, LISA will be able to provide precise sky locations months or even years before merger, allowing electromagnetic observatories to monitor for counterparts that might precede, as well as follow, the merger event.  In addition, whereas sources are likely to be on orbits with very low eccentricity once they reach the LIGO band, sources formed in highly dynamical environments like globular clusters could have sizable eccentricities in the LISA band, so observing that epoch would greatly inform our understanding of the formation mechanisms for these sources \cite{CutlerWP}.

Aside from the extremely rich astrophysics and fundamental physics that LISA can probe for massive binaries, LISA is expected to observe the infall of stellar-mass black holes into supermassive black holes, referred to as extreme mass ratio inspirals (EMRIs).  These events complement the observation of comparable-mass mergers, in that the stellar black hole serves as an effective test mass, probing the essentially static spacetime of the supermassive black hole.  In this way, EMRIs provide an equally strong-field probe, but of a system that is well understood analytically both within general relativity (GR) and within an ever-increasing number of alternative theories.  Therefore, these events can be leveraged as laboratories for placing limits on deviations from GR, or possibly even providing evidence for a theory beyond GR \cite{BerryWP}.

In addition to black-hole binaries, LISA will resolve tens of thousands of galactic binaries, primarily made up of white dwarfs, with some potential contribution from neutron stars and black holes as well.  In addition to resolvable binaries, white dwarf binaries in our galaxy are so numerous that they will provide a stochastic background.  In much the same way that resolvable and unresolvable supermassive black-hole binaries observed by PTAs will inform our understanding of their dynamics and distributions, LISA's observations will provide a tremendous advance in our understanding of compact objects in our galaxy.  Beyond the sources anticipated based on well-established empirical astrophysics, LISA could also be sensitive to more exotic sources such as primordial stochastic backgrounds from some models of inflation.  Since, unlike photons, gravitational waves travel through dense matter, plasma, strong electromagnetic fields, or anything else that might stand in their way, they are ideal probes of the very early Universe beyond the surface of last scattering.  While concordance cosmology does not predict a signature from that period that is observable by LISA, it is important to keep in mind that we truly have no other way of knowing with any degree of certainty.  Even if nothing exotic is observed by LISA in this regard, it still is clear that gravitational waves provide an ideal probe of the very early Universe, and it is likely that they will eventually better inform our understanding of this epoch \cite{LittenbergWP}.

\section*{Conclusions}

It is certainly true that no other subfield of astronomy has undergone as much development in the last decade than gravitational-wave astrophysics.  The promise of this field was clearly recognized in the past, and was primarily responsible for previous decadal endorsements.  Far more of the promise of ground-based observation has been realized at this point in time than most people would have anticipated, but it is important to keep in mind that the results so far are very much only the tip of the iceberg.  This field will eventually revolutionize our understanding of how stars of all masses and at all distances are formed and evolve, and how gravity itself operates.  All it requires of us is the foresight to invest in this field now, when the opportunities for transformational breakthroughs abound, and the possibilities seem endless.

\bibliography{LISA}

\begin{thebibliography}{38}
\providecommand{\natexlab}[1]{#1}
\providecommand{\url}[1]{\texttt{#1}}
\expandafter\ifx\csname urlstyle\endcsname\relax
  \providecommand{\doi}[1]{doi: #1}\else
  \providecommand{\doi}{doi: \begingroup \urlstyle{rm}\Url}\fi

\bibitem[{The LIGO Scientific Collaboration} et~al.(2016){The LIGO Scientific
  Collaboration}, {the Virgo Collaboration}, Abbott, et~al.]{Abbott:2016blz}
{The LIGO Scientific Collaboration}, {the Virgo Collaboration}, B.~P. Abbott,
  et~al.
\newblock {Observation of Gravitational Waves from a Binary Black Hole Merger}.
\newblock \emph{Phys. Rev. Lett.}, 116:\penalty0 061102, 2016.

\bibitem[{The LIGO Scientific Collaboration} et~al.(2017{\natexlab{a}}){The
  LIGO Scientific Collaboration}, {the Virgo Collaboration}, Abbott,
  et~al.]{TheLIGOScientific:2017qsa}
{The LIGO Scientific Collaboration}, {the Virgo Collaboration}, B.~P. Abbott,
  et~al.
\newblock {GW170817: Observation of Gravitational Waves from a Binary Neutron
  Star Inspiral}.
\newblock \emph{Phys. Rev. Lett.}, 119:\penalty0 161101, 2017{\natexlab{a}}.

\bibitem[{The LIGO Scientific Collaboration} et~al.(2017{\natexlab{b}}){The
  LIGO Scientific Collaboration}, {the Virgo Collaboration}, Abbott,
  et~al.]{GBM:2017lvd}
{The LIGO Scientific Collaboration}, {the Virgo Collaboration}, B.~P. Abbott,
  et~al.
\newblock {Multi-messenger Observations of a Binary Neutron Star Merger}.
\newblock \emph{Astrophys. J.}, 848:\penalty0 L12, 2017{\natexlab{b}}.

\bibitem[Armano et~al.(2016)]{LPF16}
M.~Armano et~al.
\newblock {Sub-Femto-$g$ Free Fall for Space-Based Gravitational Wave
  Observatories: LISA Pathfinder Results}.
\newblock \emph{Phys. Rev. Lett.}, 116:\penalty0 231101, 2016.
\newblock \doi{10.1103/PhysRevLett.116.231101}.

\bibitem[Armano et~al.(2018)]{LPF18}
M.~Armano et~al.
\newblock {Beyond the Required LISA Free-Fall Performance: New LISA Pathfinder
  Results down to $20\text{ }\text{ }\ensuremath{\mu}\mathrm{Hz}$}.
\newblock \emph{Phys. Rev. Lett.}, 120:\penalty0 061101, 2018.
\newblock \doi{10.1103/PhysRevLett.120.061101}.

\bibitem[{LIGO Scientific Collaboration}(2018)]{LSCInstrument}
{LIGO Scientific Collaboration}.
\newblock {Instrument Science White Paper 2018}, 2018.
\newblock \url{https://dcc.ligo.org/LIGO-T1800133/public}.

\bibitem[{The LIGO Scientific Collaboration} et~al.(2018){The LIGO Scientific
  Collaboration}, {the Virgo Collaboration}, Abbott, et~al.]{catalog}
{The LIGO Scientific Collaboration}, {the Virgo Collaboration}, B.~P. Abbott,
  et~al.
\newblock {GWTC-1: A Gravitational-Wave Transient Catalog of Compact Binary
  Mergers Observed by LIGO and Virgo during the First and Second Observing
  Runs}.
\newblock \emph{arXiv e-prints}, art. arXiv:1811.12907, Nov 2018.

\bibitem[Lazzarini et~al.(2016)]{ligoaplus}
A.~Lazzarini et~al.
\newblock {What Comes Next for LIGO? Planning for the post-detection era in
  gravitational-wave detectors and astrophysics}, 2016.
\newblock LIGO Document P1600350.

\bibitem[{Yu} et~al.(2018)]{ligolf}
Hang {Yu} et~al.
\newblock {Prospects for Detecting Gravitational Waves at 5 Hz with
  Ground-Based Detectors}.
\newblock \emph{Phys. Rev. Lett.}, 120:\penalty0 141102, Apr 2018.
\newblock \doi{10.1103/PhysRevLett.120.141102}.

\bibitem[{The LIGO Scientific Collaboration} et~al.(2017{\natexlab{c}}){The
  LIGO Scientific Collaboration}, {the Virgo Collaboration}, Abbott,
  et~al.]{mma}
{The LIGO Scientific Collaboration}, {the Virgo Collaboration}, B.~P. Abbott,
  et~al.
\newblock Multi-messenger observations of a binary neutron star merger.
\newblock \emph{Astrophys. J.}, 848\penalty0 (2):\penalty0 L12,
  2017{\natexlab{c}}.
\newblock \doi{10.3847/2041-8213/aa91c9}.

\bibitem[{The LIGO Scientific Collaboration} et~al.(2017{\natexlab{d}}){The
  LIGO Scientific Collaboration}, {the Virgo Collaboration}, Abbott,
  et~al.]{bns}
{The LIGO Scientific Collaboration}, {the Virgo Collaboration}, B.~P. Abbott,
  et~al.
\newblock Gw170817: Observation of gravitational waves from a binary neutron
  star inspiral.
\newblock \emph{Phys. Rev. Lett.}, 119:\penalty0 161101, 2017{\natexlab{d}}.
\newblock \doi{10.1103/PhysRevLett.119.161101}.
\newblock URL \url{https://link.aps.org/doi/10.1103/PhysRevLett.119.161101}.

\bibitem[Abbott et~al.(2017{\natexlab{a}})]{grb}
B.~P. Abbott et~al.
\newblock Gravitational waves and gamma-rays from a binary neutron star merger:
  {GW}170817 and {GRB} 170817a.
\newblock \emph{Astrophys. J.}, 848\penalty0 (2):\penalty0 L13, oct
  2017{\natexlab{a}}.
\newblock \doi{10.3847/2041-8213/aa920c}.

\bibitem[Abbott et~al.(2017{\natexlab{b}})]{hubble}
B.~P. Abbott et~al.
\newblock {A gravitational-wave standard siren measurement of the Hubble
  constant}.
\newblock \emph{Nature}, 551:\penalty0 85--88, Nov 2017{\natexlab{b}}.
\newblock \doi{10.1038/nature24471}.

\bibitem[Akrami et~al.(2018)]{Akrami:2018odb}
Y.~Akrami et~al.
\newblock {Planck 2018 results. X. Constraints on inflation}.
\newblock {arXiv:1807.06211 [astro-ph.CO]}, 2018.

\bibitem[Ade et~al.(2018)]{Ade:2018gkx}
P.~A.~R. Ade et~al.
\newblock {BICEP2 / Keck Array X: Constraints on Primordial Gravitational Waves
  using Planck, WMAP, and New BICEP2/Keck Observations through the 2015
  Season}.
\newblock \emph{Submitted to: Phys. Rev. Lett.}, 2018.

\bibitem[Kamionkowski and Kovetz(2016)]{Kamionkowski:2015yta}
Marc Kamionkowski and Ely~D. Kovetz.
\newblock {The Quest for B Modes from Inflationary Gravitational Waves}.
\newblock \emph{Ann. Rev. Astron. Astrophys.}, 54:\penalty0 227--269, 2016.
\newblock \doi{10.1146/annurev-astro-081915-023433}.

\bibitem[Ahmed et~al.(2014)]{Ahmed:2014ixy}
Z.~Ahmed et~al.
\newblock {BICEP3: a 95GHz refracting telescope for degree-scale CMB
  polarization}.
\newblock \emph{Proc. SPIE Int. Soc. Opt. Eng.}, 9153:\penalty0 91531N, 2014.
\newblock \doi{10.1117/12.2057224}.

\bibitem[Essinger-Hileman et~al.(2014)]{Essinger-Hileman:2014pja}
Thomas Essinger-Hileman et~al.
\newblock {CLASS: The Cosmology Large Angular Scale Surveyor}.
\newblock \emph{Proc. SPIE Int. Soc. Opt. Eng.}, 9153:\penalty0 91531I, 2014.
\newblock \doi{10.1117/12.2056701}.

\bibitem[Arnold et~al.(2014)]{Arnold:2014qym}
K.~Arnold et~al.
\newblock {The Simons Array: expanding POLARBEAR to three multi-chroic
  telescopes}.
\newblock \emph{Proc. SPIE Int. Soc. Opt. Eng.}, 9153:\penalty0 91531F, 2014.
\newblock \doi{10.1117/12.2057332}.

\bibitem[Aguirre et~al.(2018)]{Ade:2018sbj}
James Aguirre et~al.
\newblock {The Simons Observatory: Science goals and forecasts}.
\newblock {arXiv:1808.07445 [astro-ph.CO]}, 2018.

\bibitem[Abazajian et~al.(2016)]{Abazajian:2016yjj}
Kevork~N. Abazajian et~al.
\newblock {CMB-S4 Science Book, First Edition}.
\newblock {arXiv:1610.02743 [astro-ph.CO]}, 2016.

\bibitem[Gualtieri et~al.(2017)]{Gualtieri:2017zcz}
R.~Gualtieri et~al.
\newblock {SPIDER: CMB Polarimetry from the Edge of Space}.
\newblock In \emph{{17th International Workshop on Low Temperature Detectors
  (LTD 17) Kurume City, Japan, July 17-21, 2017}}, 2017.
\newblock URL
  \url{http://lss.fnal.gov/archive/2017/conf/fermilab-conf-17-570-ae.pdf}.

\bibitem[Gandilo et~al.(2016)]{Gandilo:2016sqn}
Natalie~N. Gandilo et~al.
\newblock {The Primordial Inflation Polarization Explorer (PIPER)}.
\newblock \emph{Proc. SPIE Int. Soc. Opt. Eng.}, 9914:\penalty0 99141J, 2016.
\newblock \doi{10.1117/12.2231109}.

\bibitem[Suzuki et~al.(2018)]{Suzuki:2018cuy}
A.~Suzuki et~al.
\newblock {The LiteBIRD Satellite Mission - Sub-Kelvin Instrument}.
\newblock In \emph{{17th International Workshop on Low Temperature Detectors
  (LTD 17) Kurume City, Japan, July 17-21, 2017}}, 2018.
\newblock \doi{10.1007/s10909-018-1947-7}.

\bibitem[Young et~al.(2018)]{Young:2018aby}
Karl Young et~al.
\newblock {Optical Design of PICO, a Concept for a Space Mission to Probe
  Inflation and Cosmic Origins}.
\newblock {arXiv:1808.01369 [astro-ph.IM]}, 2018.

\bibitem[{Arzoumanian} et~al.(2018)]{NANOGrav}
Z.~{Arzoumanian} et~al.
\newblock {The NANOGrav 11 Year Data Set: Pulsar-timing Constraints on the
  Stochastic Gravitational-wave Background}.
\newblock \emph{Astrophys. J.}, 859:\penalty0 47, 2018.
\newblock \doi{10.3847/1538-4357/aabd3b}.

\bibitem[{Shannon} et~al.(2015)]{Parkes}
R.~M. {Shannon} et~al.
\newblock {Gravitational waves from binary supermassive black holes missing in
  pulsar observations}.
\newblock \emph{Science}, 349:\penalty0 1522--1525, 2015.
\newblock \doi{10.1126/science.aab1910}.

\bibitem[{NANOGrav Collaboration} et~al.(2018{\natexlab{a}}){NANOGrav
  Collaboration}, {Arzoumanian}, et~al.]{nano}
{NANOGrav Collaboration}, Z.~{Arzoumanian}, et~al.
\newblock {The NANOGrav 11 Year Data Set: Pulsar-timing Constraints on the
  Stochastic Gravitational-wave Background}.
\newblock \emph{Astrophys. J.}, 859:\penalty0 47, 2018{\natexlab{a}}.
\newblock \doi{10.3847/1538-4357/aabd3b}.

\bibitem[{NANOGrav Collaboration} et~al.(2016){NANOGrav Collaboration},
  {Arzoumanian}, et~al.]{nanospec}
{NANOGrav Collaboration}, Z.~{Arzoumanian}, et~al.
\newblock {The NANOGrav Nine-year Data Set: Limits on the Isotropic Stochastic
  Gravitational Wave Background}.
\newblock \emph{Astrophys. J.}, 821:\penalty0 13, 2016.
\newblock \doi{10.3847/0004-637X/821/1/13}.

\bibitem[{NANOGrav Collaboration} et~al.(2018{\natexlab{b}}){NANOGrav
  Collaboration}, {Aggarwal}, et~al.]{nanocw}
{NANOGrav Collaboration}, K.~{Aggarwal}, et~al.
\newblock {The NANOGrav 11-Year Data Set: Limits on Gravitational Waves from
  Individual Supermassive Black Hole Binaries}.
\newblock \emph{arXiv e-prints}, art. arXiv:1812.11585, 2018{\natexlab{b}}.

\bibitem[{NANOGrav Collaboration} et~al.(2015){NANOGrav Collaboration},
  {Arzoumanian}, et~al.]{nanobwm}
{NANOGrav Collaboration}, Z.~{Arzoumanian}, et~al.
\newblock {NANOGrav Constraints on Gravitational Wave Bursts with Memory}.
\newblock \emph{Astrophys. J.}, 810:\penalty0 150, 2015.
\newblock \doi{10.1088/0004-637X/810/2/150}.

\bibitem[on~the Review of Progress Toward~the Decadal~Survey(2016)]{NWNH}
Committee on~the Review of Progress Toward~the Decadal~Survey.
\newblock \emph{{New Worlds, New Horizons: A Midterm Assessment}}.
\newblock The National Academies Press, Washington, DC, 2016.
\newblock \doi{10.17226/23560}.

\bibitem[Berti et~al.(2019)Berti, Shoemaker, Barausse, Cholis,
  Garc\'ia-Bellidoc, Hughes, Kelly, Kovetz, Littenberg, Livas, Schnittman, and
  Yunes]{BertiWP}
Emanuele Berti, Deirdre Shoemaker, Enrico Barausse, Ilias Cholis, Juan
  Garc\'ia-Bellidoc, Scott~A. Hughes, Bernard Kelly, Ely~D. Kovetz, Tyson~B.
  Littenberg, Jeffrey Livas, Jeremy Schnittman, and Nicolas Yunes.
\newblock Astro2020 science white paper: Tests of general relativity and
  fundamental physics with space-based gravitational wave detectors.
\newblock \emph{submission to the 2020-2030 Astronomy and Astrophysics Decadal
  Survey (Astro2020)}, 2019.

\bibitem[Colpi et~al.(2019)Colpi, Holley-Bockelmann, Bogdanovi\'c, Natarajan,
  Sesana, Tremmel, Comerford, Barausse, Berti, Volonteri, Khan, and
  McWilliams]{ColpiWP}
M.~Colpi, K.~Holley-Bockelmann, T.~Bogdanovi\'c, P.~Natarajan, A.~Sesana,
  M.~Tremmel, J.~Comerford, E.~Barausse, E.~Berti, M.~Volonteri, F.~M. Khan,
  and S.~T. McWilliams.
\newblock The {GW} view of massive black holes.
\newblock \emph{submission to the 2020-2030 Astronomy and Astrophysics Decadal
  Survey (Astro2020)}, 2019.

\bibitem[Sesana(2016)]{Sesana:2016ljz}
Alberto Sesana.
\newblock {Prospects for Multiband Gravitational-Wave Astronomy after
  GW150914}.
\newblock \emph{Phys. Rev. Lett.}, 116:\penalty0 231102, 2016.
\newblock \doi{10.1103/PhysRevLett.116.231102}.

\bibitem[Cutler et~al.(2019)Cutler, Berti, Jani, Kovetz, Littenberg, Randall,
  Vitale, and Wong]{CutlerWP}
Curt Cutler, Emanuele Berti, Karan Jani, Ely Kovetz, Tyson Littenberg, Lisa
  Randall, Salvatore Vitale, and Kaze W.~K. Wong.
\newblock What we can learn from multi-band gravitational-wave observations of
  black hole binaries.
\newblock \emph{submission to the 2020-2030 Astronomy and Astrophysics Decadal
  Survey (Astro2020)}, 2019.

\bibitem[Berry et~al.(2019)Berry, Hughes, Sopuerta, Chua, Heffernan, Miller,
  and Sesana]{BerryWP}
Christopher P.~L. Berry, Scott~A. Hughes, Carlos~F. Sopuerta, Alvin J.~K. Chua,
  Anna Heffernan, M.~Coleman Miller, and Alberto Sesana.
\newblock The unique potential of extreme mass-ratio inspirals for
  gravitational-wave astronomy.
\newblock \emph{submission to the 2020-2030 Astronomy and Astrophysics Decadal
  Survey (Astro2020)}, 2019.

\bibitem[Littenberg et~al.(2019)Littenberg, Breivik, Brown, Eracleous, Hermes,
  Kremer, Kupfer, and Larson]{LittenbergWP}
Tyson~B. Littenberg, Katelyn Breivik, Warren~R. Brown, Michael Eracleous, J.~J.
  Hermes, Kyle Kremer, Thomas Kupfer, and Shane~L. Larson.
\newblock Gravitational wave survey of galactic ultra compact binaries.
\newblock \emph{submission to the 2020-2030 Astronomy and Astrophysics Decadal
  Survey (Astro2020)}, 2019.

\end{thebibliography}

\end{document}